# Minimal effect of prescribed burning on fire spread rate and intensity in savanna ecosystems

## Aristides Moustakas[1,*] & Orestis Davlias[2]

1. Natural History Museum of Crete, University of Crete, Heraklion, Greece
2. School of Electrical and Electronic Engineering, Technical University of Crete, Greece

* Corresponding author
Email: arismoustakas@gmail.com

**Abstract**

Fire has been an integral part of the Earth for millennia. Several recent wildfires have exhibited an unprecedented spatial and temporal extent and their control is beyond national firefighting capabilities. Prescribed or controlled burning treatments are debated as a potential measure for ameliorating the spread and intensity of wildfires. Machine learning analysis using random forests was performed in a spatio-temporal data set comprising a large number of savanna fires across 22 years. Results indicate that fire return interval was not an important predictor of fire spread rate or fire intensity, having a feature importance of 3.5%, among eight other predictor variables. Manipulating burn seasonality showed a feature importance of 6% or less regarding fire spread rate or fire intensity. While manipulated fire return interval and seasonality moderated both fire spread rate and intensity, their overall effects were low in comparison with meteorological (hydrological and climatic) variables. The variables with the highest feature importance regarding fire spread rate resulted in fuel moisture with 21%, relative humidity with 15%, wind speed with 14%, and last years' rainfall with 14%. The variables with the highest feature importance regarding
fire intensity included fuel load with 21.5%, fuel moisture with 16.5%, relative humidity with 12.5%, air temperature with 12.5%, and rainfall with 12.5%, Predicting fire spread rate and intensity has been a poor endeavour thus far and we show that more data of the variables already monitored would not result in higher predictive accuracy.

**Introduction**

Humans are the sole species that can proactively ignite or suppress fires, shaping landscapes (Brown et al. 2009). Fire is not necessarily a hazard; it is an indissoluble ecosystem component having also several positive ecosystem aspects (Lino et al. 2019). However, recent wildfires are extended to very large surface areas, and their duration is exhibiting longer temporal scales. Savannas cover about a fifth of the global land surface and about half of the area of Africa, Australia, and South America (Scholes and Archer 1997, Moustakas et al. 2010) and these areas tend to burn frequently and account for a large number of the world's fires and fire emissions (Randerson et al. 2018).

Despite several decades of research our understanding of factors shaping fire dynamics remains poor. The causes of increased fire risk are linked to climate change, resulting in increased prevalence of hotter and drier periods (Di Virgilio et al. 2019, Madadgar et al. 2020). However, increased fire risk can also be driven by economic and social changes, and political decisions (Hesseln 2000, Tàbara et al. 2003, Poudyal et al. 2012, Qayum et al. 2020). In some areas fires are exacerbated by rural land abandonment and increased ignition probability in the growing rural-to-urban interface (Flannigan et al. 2009, Nishino 2019). In some other areas anthropogenic fire, at a higher frequency than the natural fire return interval, has traditionally been practiced for pasture, agriculture or fire prevention (McDaniel et al. 2005). Increased fire risk merits research, regarding how fire spreads in these complex fuel loads and what is the relative contribution of each factor into fire spread or intensity (Flannigan et al. 2009).

It is debated whether increased fire risk can be moderated by prescribed (also referred to as controlled) burning (Penman et al. 2011). By human-induced controlled burning there is a potential for reducing the fuel load for large wildfires to occur (North et al. 2012). This can modulate both fire spread and intensity (Volkova and Weston 2019). This is feasible by burning during seasons that fuel is more humid or by burning frequently so that fuel load is lower (Higgins et al. 2007).

Manipulating fire frequency and seasonality has potential effects on carbon emissions (Van der Werf et al. 2003), soil (Francos et al. 2019), carbon balance (Merino et al. 2019), species composition (Fisher et al. 2009), water & air quality (Lucas-Borja et al. 2019, Ravi et al. 2019) and thus the overall trade-offs between frequent controlled fire, fire suppression, or fire control after ignition need to be accounted for (Tilman et al. 2000). More frequent burning may increase carbon emissions (Brown and Johnstone 2011) in comparison to the ones that would have occurred in the absence of human intervention potentially feeding back the climate-fire loop. On the other hand, fire suppression can result in increased carbon uptake by land (Arora and Melton 2018, Jones et al. 2019) as well as it may play an important role in offsetting impacts of increased ignitions (Keeley et al. 1999). In addition, more frequent burning shapes the environment by potentially selecting for fire prone species, or fast growing plant species (Hengst and Dawson 1994, Schwilk and Ackerly 2001). Fire may also facilitate invasive alien species (Fisher et al. 2009). Fire spread and intensity apart from humans depends also on biological, meteorological, and physical factors. The inclusion of meteorological ecosystem variables can enhance the understanding of fire spread rate (Taufik et al. 2017). Other studies have indicated that vapor pressure deficit (a variable based on temperature and relative humidity) on can be the best meteorological variable for fire prediction on many timescales, as it scales linearly with evapotranspiration and reflects fine fuel moisture well (Chen et al. 2020). There is species adaptation to fire (Schwilk and Ackerly 2001) by partitioning underground resources and thus including species identity could increase the predictive ability. Enhanced nutrients after fire (Van de Vijver et al. 1999), or drought-related indexes such as the standardized precipitation index (Keyantash 2018) or factors influencing fuel load are likely to improve the predictive accuracy.

There are several studies regarding the effect of prescribed burning on fire spread and intensity. Several other studies quantify fire spread and intensity using additional climatic, hydrological, or biological explanatory variables (Espinosa et al. 2019, Lucas-Borja et al. 2019). While these studies strongly facilitate our understanding on fire dynamics, common hindrances to generalizing or training predictive models include the limited sample size of fire events (Ching et al. 2018, Nikolopoulos et al. 2018). Experimental burning treatments often have limited duration due to associated costs, legal limitations, or ethics resulting in limited spatial or temporal sample size replicate. Naturally occurring fires (i.e. not controlled fire experiments) provide data for understanding how fire spread or intensity is influenced by hydrological, climatic, and biological variables but they do not provide the what-if manipulative experimental option of quantifying fire properties had controlled burning or fire suppression preceded the fire event.

In this study we quantify the independent effect of human-induced controlled burning, meteorological (climatic & hydrological), and geological characteristics on savanna fire spread rate and intensity across a large number of fires. The prescribed burning return interval expands both below and above the long-term mean "natural" return interval in the absence of prescribed burning, thereby including both more frequent burning as well as fire suppression. We explicitly refrained from formulating any hypotheses regarding the effects of prescribed burning as often when we make a hypothesis, we may become attached to it (Platt 1964); instead we performed data-driven analysis, making the implicit hypothesis that an underlying dependence between collected data can be objectively mined (Moustakas et al. 2019). Such approaches (van Helden 2013) are not competitive to hypothesis-led studies; instead they are complementary and iterative with them (Kell and Oliver 2004).

**Methods**

A 22-year spatio-temporal dataset comprised of 956 savanna fires in the Kruger National Park (KNP), South Africa (Fig. 1a) was used and is fully described in (Govender et al. 2006). All data used here have been published elsewhere addressing different questions (Govender et al. 2006, Higgins et al. 2007). KNP is located on a low-lying savanna landscape covering at total area of 19,485 km² (Fig. 1b), and it forms one of the best studied ecosystems in terms of fire (Higgins et al. 2007). KNP is comprised of four savanna landscapes with two distinctive soil bedrocks: granites Pretoriuskop (long-term Mean Annual Precipitation (MAP) = 737 mm) and Skukuza (MAP = 550 mm), and basalts: Satara (MAP = 544 mm) and Mopane (MAP = 496 mm). (Govender et al. 2006). Within the park there are long-term Experimental Burning Plots (EBPs) where fire is manipulated as a treatment. Within each EBP experimental burns regarding fire frequency (N=5 treatments) in combination with fire seasonality (N=4 treatments) are conducted as a split-block randomized design experiment where each EBP consists of 12 of the possible treatments including a control treatment i.e. no fire (Fig. 1c). The experiment is replicated four times within each of the four landscapes of the park resulting in 16 EBPs. All fire data employed here occur exclusively in 7 ha surface area treatments . Thus the experimental area on each EBP is 12 treatments * 7 ha=84 ha, while the total experimental area is 84 ha/EBP * 16 EBPs = 1344 ha.

*Explanatory variables*

In the analysis conducted here, nine predictor/explanatory/independent variables were employed spanning prescribed fire characteristics, fuels, soil, and meteorology. These included: fire return interval, fuel moisture, fuel load, air temperature, relative humidity, wind speed, soil, rainfall, soil, and burn season. A detailed data description is provided in (Govender et al. 2006). All data were collected by the KNP scientific services. In brief: in terms of spatial extent each experimental burning treatment (i.e. fire event) consisted of a seven hectares surface area burn plot. Each of the 956 fire events consists of a combination of a fire return interval and fire seasonality. A unique fire frequency and seasonality treatment is applied on each treatment through years. Prescribed burning included a manipulated fire return interval of 1, 2, 3, 4, and 6 years, when the natural fire return interval is 4.5 years. Fire seasonality as prescribed burning intervention included autumn, winter, spring, and summer burns.

Fuel load before each fire event was estimated in the field (at a temporal scale of approximately an hour before each fire) at the spatial scale of each 7-ha treatment within each EBP using a disc pasture meter (Bransby and Tainton 1977) calibrated for KNP (Trollope and Potgieter 1986). An average of 100 spatial random records of fuel load within each treatment prior to each fire event in kg m$^{-2}$ was used as a fuel load value for the fire event.

The predominant soil bedrock on each experimental burning treatment was recorded at the scale of each EBP. These included basalt or granite depending on the landscape of the EBP.

Prior to each fire, the moisture content of the fuel load was estimated by sampling four 100-g swards at the scale of each treatment within each EBP. The samples were stored in air-proof bottles and sequentially dried moisture content was expressed as a percentage of dry mass:

$$M = [(W - D)/D] * 100 \quad [1]$$

where $M$ is the fuel moisture content, $W$ the wet mass of the grass sward sample and $D$ the dry mass of the sample.

Rainfall was quantified as the mean precipitation over the last 24 months (in mm year$^{-1}$) before each fire event (two hydrological years not calendar years) at the scale of each EBP using a permanent gauge. The mean of two years instead of one was used because the effect of rainfall on perennial grasses persists for more than a year (Van Wilgen et al. 2004). The mean wind speed recorded in m sec$^{-1}$, air temperature in $^{\circ}$C, and the relative humidity (%) of the air during each fire event were also retrieved at the scale of each EBP using the permanent EBP weather recording gauges. These weather data were averaged over the duration of each fire event.

*Fire spread rate and fire intensity*

The dependent variables analyzed included fire spread rate and fire intensity; the analysis was repeated twice, once for each dependent variable, all else been equal regarding the explanatory variables.

Fire spread rate and fire intensity were quantified as a function of fire return interval, fire seasonality, wind speed, fuel load, soil type, relative humidity, fuel moisture, air temperature, and rainfall during the last two years (explanatory variables). These are variables commonly used for monitoring fires. The study plots are generally flat and for this reason no topographical variables were used

Fire spread rate was estimated as:

$$r = A/(L * T) \quad [2]$$

where $r$ is the rate of spread (m sec$^{-1}$), $A$ is the area burned as a head fire (m$^2$), $L$ the mean fire front length (m) and $T$ the duration of the fire (s); (Govender et al. 2006). All fire spread rate values were ranked from lowest to highest and binned to five bin classes with approximately equal data representation within each bin. Bin one represents the slower spreading fires while bin five the fastest. Bins instead of the original data values were used due to the high level of variance in the original independent variables. A machine learning derived prediction will be classified as error unless the predicted value is identical with the original value even if the prediction is very close to the actual original value. Techniques to standardize, normalize, and/or adjust the original data prior to the analysis were explored but the actual RF output yielded predictions of < 20% accuracy. We therefore proceeded with binning fire spread rate and using five bins throughout the analysis.

Fire intensity was estimated as Byram's fire-line intensity:

$I = H * w * r$ [3]

where $I$ is fire intensity (kW m$^{-1}$), $H$ is heat yield (kJ g$^{-1}$), $w$ is the combusted fuel mass (g m$^{-2}$), and $r$ is the rate of spread of the heat fire front (m sec$^{-1}$); (Byram 1959). Heat of combustion was determined with a calorimeter, and values were corrected to heat yields to allow for incomplete combustion in vegetation fires (Byram 1959, Govender et al. 2006). All fire intensity values were ranked from lowest to highest and binned to five bin classes from bin class one representing the lowest fire spread rate to bin class five representing the highest fire spread rate. Approximately equal data representation per bin class was used. Our rationale behind binning fire intensity is identical to the one explained in the above paragraph regarding binning fire spread rate.

*Data analytics*

Random Forests (RF); (Breiman 2001) employ boosting (Schapire et al. 1998) and bagging algorithms (bootstrap aggregating; (Breiman 1996)) of the Classification And Regression Tree (CART; (Breiman 2017)) model, and use a more random but more efficient node splitting strategy than standard CARTs (Liaw and Wiener 2002). RFs have been employed in a wide variety of classification and prediction problems (Daliakopoulos et al. 2017, San-Miguel et al. 2020, Zhu et al. 2020) as they rank among the most efficient analytic tools for extracting information in noisy, complex, and high dimensional datasets (Wager et al. 2014, Scornet et al. 2015). An extensive data-driven model inter-comparison by (Fernández-Delgado et al. 2014) showed that they may be one of the strongest data analytics tools for real-world problems.

We trained and tested a RF classifier, using 10 different random states, for each train-to-testing data split candidate. The data split partitioning explored included the following training to testing data partitions: 90-10, 80-20, 75-25, 70-30, 50-50 measuring each time the model accuracy. The optimal partitioning was chosen with respect to maximizing the predictive accuracy based on selecting each time two different random sub-groups of the dataset as training and test sets for each of the different partitions. The predictive accuracy is defined as the number of correct bin classifications divided by the total number of fire instances in the dataset. The optimal train-to-testing data split is selected based on the mean predictive accuracy, as well as, the standard deviation across the trials (James et al. 2013).

Having detected the optimal training-to-testing ratio, we quantified the partial dependence (Friedman 2001, Hastie et al. 2009) between (1) fire spread rate and (2) fire intensity and two of our independent variables: fire return interval and seasonality (Pedregosa et al. 2011); (https://scikit-learn.org/stable/modules/partial_dependence.html). Partial effects in RF quantify the effect of an independent variable on the dependent variable as a percentage all else been equal (Friedman 2001, Hastie et al. 2009). For example a partial effect change of 0.05 regarding fire spread rate when the fire return interval increases from 2 years to 3 years, indicates that when fire return interval increases by a unit of one (year), there is a 5% chance of fire spread rate changing from current bin to a different bin (Zhao and Hastie 2019).

In order to quantify the importance of each of the nine explanatory variables (features) we calculated the mean decrease in impurity for each variable (Pedregosa et al. 2011). We used the sklearn's embedded feature importance method, that evaluates the impurity importance based feature importances of the forest, along with their inter-trees variability (Pedregosa et al. 2011). The feature importance method used tends to be more biased towards high cardinality features and thus the numerical features' importance (Fuel moisture, Fuel load, etc.) may be boosted over the categorical ones (Season and Soil); (Pedregosa et al. 2011). So, in order to avoid any misleading results we also examined the target response (the probability of each bin-class) of each of the explanatory variables on each of the dependent variables. More specifically for each of the target variables (explanatory variables) $X_S$ and their complements $X_C$ we calculate the integral:

$$pd_{X_S}(x_S) \stackrel{def}{=} \mathbb{E}_{X_C}[f(x_S, X_C)] = \int f(x_S, x_C)p(x_C)dx_C,$$

[4]

where f($x_S$, $x_C$) is the response function for a given sample whose values are defined by $x_S$ for the features in Xs, and by $x_C$ for each feature $X_C$. The integrals were calculated using the recursive sklearn's method plot_partial_dependence (Pedregosa et al. 2011); (https://scikit-learn.org/stable/auto_examples/ensemble/plot_forest_importances.html). The standard deviation across the 10 random states was calculated In order to quantify 95% confidence intervals of the effect of each feature importance.

The code used for implementing all random forests analytics and visualization in Python is provided in the supplementary material.

Linear regression between fuel load (target variable) and fire return interval was performed to quantify whether a linear relationship between fuel load and fire return interval exists. We sought to quantify the potential correlation between the two prescribed burning variables (years since last fire & fire seasonality) and fuel load as more frequent wet season fires may result in less fuel load and thus the three independent variables may exhibit interaction effects. Analysis of variance (ANOVA) was performed between fuel load (target variable) and fire return interval as a continuous variable or fire seasonality as a factor. This was conducted in order to quantify the relationship between years since last fire as well as fire season on fuel load accumulation. ANOVA results were plotted at each factor level mean, the overall mean, and the 95% decision limits (Schilling 1973). If a point falls outside the decision limits, then evidence exists that the factor level mean represented by that point is significantly different (at a 5% significant level) from the overall mean (Schilling 1973).

**Results**

Fires were not evenly distributed between years with some years exhibiting higher activity than others (Fig. 1d). However there were only two years (1984 and 1985) without any fires, and even years with low number of fires included at least 5 fires (Fig. 4d). The majority of fires had a fire return interval of two or three years, while fewer fires had a fire return interval of four or six years (Fig. 1e). The majority of fires included winter fires, followed by summer fires (Fig. 1f). Spring and autumn fires included at least 180 fires each (Fig. 1f).

The optimal training-to-testing data split minimizing mean error included a 75-25 % partition for fire spread rate and 70 – 30 % for fire intensity.

As there are five bins for fire spread rate, five partial effect plots are needed to visualize the effect of fire return interval and five for the effect of fire season on fire spread rate (one for each bin size). For economy of space the partial effects of the first bin size for fire return interval and fire season are plotted on fire spread rate, while the full plots across all bins are plotted in the supplementary material. On the first bin size of fire spread rate, the chance of current bin changing to another bin increased by a small fraction between a 1-year return interval and a 6-year return interval; there was an increase from 0.19 when annual burn occurred to 0.235 of a bin when the fire return interval was six years (Fig. 2a). The chance of changing fire spread rate bin was minimized during winter months (partial dependence value 0.14), increased with an almost identical effect of spring and autumn fires spread rate (partial dependence close to 0.23), while it maximized during summer months (partial dependence 0.24); (Fig. 2b). In all other bins the partial effects of fire return interval and fire season on fire spread rate remained very small; the chance of bin change rarely exceeded 0.25 and in general was unaffected by changes in values of fire return interval (Supplementary Fig. 1) and fire season (Supplementary Fig. 2).

As there are five bins for fire intensity, five partial effect plots are needed to visualize the effect of fire return interval and five for the effect of fire season on fire spread intensity (one for each bin size). For economy of space the partial effects of the first bin size for fire return interval and fire season are plotted on fire intensity, while the full plots across all bins are plotted in the supplementary material. On the first bin size of fire intensity, the chance of current bin changing to another bin increased by a small fraction between a 1-year return interval and a 6-year return interval; values of fire return interval bin increased from 0.19 when annual burn occurred to 0.245 when the fire return interval was six years (Fig. 2c). The chance of the first fire intensity bin changing to another was minimized during winter months (partial dependence of 0.15), increased with spring fires to value of 0.21 and autumn to 0.22 with summer fires exhibiting the highest value of 0.225 (Fig. 2d). In all other bins the partial effects of fire return interval and fire season on fire intensity remained very small and in general was unaffected by changes in values of fire return interval (Supplementary Fig. 3) and fire season (Supplementary Fig. 4).

Feature importance regarding fire spread rate resulted in (by reducing feature importance) fuel moisture with a feature importance of 21%, followed by relative humidity with 15%, wind speed 14%, last years' rainfall 14%, fuel load 13%, and air temperature 11.5% (Fig. 3a). Fire return interval featured an importance of 3.5% of the total (mean value) reaching a maximum feature importance of 5% (upper level of the error bar); (Fig. 3a). Seasonality of fire had a feature importance of 6% reaching a maximum of 10% (upper level of the error bar); (Fig. 3a). Soil type had a feature importance of 2% (Fig. 3a).

Regarding fire intensity fuel load had a feature importance of 21.5%, fuel moisture 16.5%, relative humidity 12.5%, air temperature 12.5%, rainfall 12.5%, and wind speed 12% (Fig. 3b). Seasonality of fire had a feature importance of 5% with a maximum of 6.5% (upper level of the error bar), while fire return interval 3.5% with a maximum of 5% (Fig. 3b). Soil type had a feature importance of 3% (Fig. 3b).

Predicting the bin size of fire spread rate with the RF trained model had a mean value of 40.5% when the train-to-testing data ratio was 50-50% with a maximum of 42.5% (upper level of the error bar) (Fig. 4a). This result was not sensitive to the overall amount of available data, as partitioning the dataset into a range of 50% training 50% testing or 90% learning 10% testing would yield very small differences in the RF model output regarding fire spread rate; the 90% learning 10% testing data partitioning ratio resulted in a mean value of 40.7% with a maximum of 47%, and thus increased the variance but not the mean accuracy (Fig. 4a). In addition, none of the data partitions (50-50% to 90-10%) resulted in significantly different accuracies, as the 95% confidence intervals of all train-to-testing data partitions are overlapping (Fig. 4a).

Predicting the bin size of fire intensity with the RF trained model spanned from 42% when the train-to-testing data ratio was 50-50% with a maximum of 44.5% (upper level of the error bar) (Fig. 4b).

This result was not sensitive to the overall amount of available data, as partitioning the dataset into a range of 50% training 50% testing or 90% learning 10% testing would yield very small differences in the RF model output predictive accuracy regarding fire intensity; the 90% learning 10% testing data partitioning ratio resulted in a mean value of 44% with a maximum of 47.5% (Fig. 4b). In addition, none of the data partitions (50-50% to 90-10%) resulted in significantly different accuracies, as the 95% confidence intervals of all train-to-testing data partitions are overlapping (Fig. 4b).

Fuel load is only weakly correlated with fire return interval with an $R^2$ = 1.9%; linear regression *Fuel load* = 3227 + 241.9 *Frequency*, SS = 44953185, F = 19.5, p << 0.001. ANOVA between fuel load and fire return interval indicated that annual burning decreased the mean fuel load as well as burning every six years (Fig. 5a). However, burning every three years increased the mean fuel load, while burning every two or four years had an effect that could not be differentiated from random (Fig. 5a). ANOVA between fuel load and seasonality of burning indicated increased fuel loads for autumn and summer burns, decreased fuel loads for winter burns, and no significant effect of spring burns (Fig. 5b).

### Discussion

The analysis performed here indicates that savanna fire spread rate is only mildly influenced by fire return interval and fire seasonality. Instead, fire spread rate is mainly determined by meterological factors; the top three most feature importance variables for fire spread rate include hydrology-related phenomena, while the next two include temperature and wind characteristics. Therefore, abrupt climatic changes regarding drought, heat, and intensive wind are highly likely to influence/increase fire spread rate. Abrupt events have also been shown to increase temporal synchrony across increasing spatial scales (i.e. generating cycles that can last longer across greater spatial extend) and to that end fires may become more temporally synchronized (Hansen et al. 2013, Sheppard et al. 2015, Moustakas et al. 2018). Based on the results derived here and despite an overall small effect, manipulating seasonality has an overall higher impact on fire spread rate and intensity than manipulating fire return interval. This means has implications for potential de-synchronization of available fuel loads by seasonally asynchronous burning in different locations thereby reducing the risks of climate-induced synchronized fires across landscapes or regions.

Fuel load had the highest feature importance for fire intensity. What is the reciprocal effect of prescribed burning on fuel load? Annual burning and winter burning reduced fuel loads. However, the overall relationship between fire return interval and fuel load is not linear (six year burning reduced mean fuel loads but two, three and four-year burns did not), and fuel load is developed rapidly since last fire (Pausas and Ribeiro 2013). Thus, frequent burning reduces fuel load but the overall effect of fire frequency on fuel load remains complex indicating that there are other interacting factors shaping fuel load. It is important to note that despite fuel load featuring first regarding fire intensity, fuel moisture and relative humidity (the $2^{nd}$ and $3^{rd}$ feature importance variables) are having a cumulative feature importance of ≈30%.

The fundamental problem of statistically producing forecasts by exploiting information from their past values only i.e. forecasting is of traditional interest to environmental science (Papacharalampous et al. 2019, Zhu et al. 2020). Predicting fire intensity or fire spread rate has in general been a poor endeavour so far; accuracy of predictions has not yet exceeded 50% (Cruz et al. 2018, Coffield et al. 2019). Note that the accuracy of predictions depends on the number of bins of fire spread rate or fire intensity where more often than not smaller number of bins result in higher number of correct predictions; here five bins are used while (Coffield et al. 2019) used three bins. The main issue identified here is that using one of the best fire datasets, in one of the best studied ecosystems on Earth regarding fire, the predictive ability of fire spread rate or fire intensity remains low (< 50%).

Interestingly, having additional data would not result in better predictions as reported here. We need additional variables data (*sensu* more different variables) not necessarily more data (*sensu* more entries of the same variables). While increased sample size enhances model training, a trained model may not go beyond the information content of the variables sampled (Evans et al. 2014, Lonergan 2014). More often than not, the more data available the more information exists, and thus the deviance between model fit and the reality as expressed from data will be minimized – but see also (Boivin and Ng 2006) and discussion in (Moustakas 2017). Recent advances in data analytics and data science bring an additional element which is to quantify the predictive ability of a trained model and its transferability of predictions in a dataset different than the one used for model training (Padarian et al. 2019). However, forecasting using a trained fire savanna model in another savanna than the one trained is not meaningful when the model accuracy is low. Thus, there is a need for developing "essential fire variables" as suggested for biodiversity or forestry (Evans and Moustakas 2016, Jetz et al. 2019) for increasing fire spread rate and intensity predictive accuracy.

While fire return interval or seasonality had a relatively small effect on ameliorating fire spread rate and intensity, intensity was best featured by fuel load. Albeit fuel loads exhibit a non-linear relationship with fire return interval, annual burn and winter fires reduced fuel loads in the savanna studied. Thus, in cases where keeping open landscapes is desired, or for agricultural purposes, or for avoiding the dominance of thorny woody species, frequent burning will reduce fuel loads as traditionally practiced through centuries (Sheuyange et al. 2005). However, prescribed burning seems unlikely to be an efficient option for controlling the spread of wildfires at least in fire-prone savannas such as the one studied. Savannas have been fire-prone ecosystems for very long time scales and fires are both natural and anthropogenic to a point that is hard to distinguish between them (O'Connor et al. 2011, Bowman et al. 2016).

It is important to note that the feature importance values regarding the effects of fire return interval or fire seasonality are relative to other variables and are affected by how many other variables are employed in the analysis. This study spans over 22 years and covering a relative large number of fires; it contains both data of fires more frequent than the long-term "natural" fire return interval as well as data on fire suppression i.e. fires with fire return interval longer than the "natural" one. However, the study length does not allow a large replicate of five or six year burns, and thus the data regarding fire suppression are under-represented. In addition, the experimental design of prescribed burning as examined here included manipulating fire season and fire return interval. However, the concept of prescribed burning is much more complex than the one examined here in the size, duration, location, method of prescribed burning. All fire events analyzed here occur in treatments of a surface area of 7 ha each. While fire spread rate and intensity are accounting for surface area, these values may be biased by the confinement of space in the experimental design. Some of factors influencing the outcome of prescribed burning unaccounted for here include the size and duration of the burns, the species identity, or the type of vegetation, the size of plant individuals (Andersen et al. 1998, Baeza et al. 2002, Moustakas 2015, Kochanski et al. 2018, Ng et al. 2020).

Prescribed burning often has more effectiveness at mitigating smaller or less intense wildfires as opposed to larger or more severe ones and fuel accumulation can limit prescribed burning effectiveness to a short after fire period (Fernandes and Botelho 2003). Prescribed burns have little-to-no effect on reducing the spread rate and intensity of fires in savanna ecosystems - this can be very different from forest ecosystems, which are overstocked with fuel, leading to more extreme fires (García-Llamas et al. 2019) – but see also (Bessie and Johnson 1995). Probably this is because savannas are mostly grasses and they recover annually. If that is the case, then savanna burning frequency can be discussed regarding mitigating emissions (Watts and Samburova 2020).

**Acknowledgements**

We thank Petros Lymberakis and Dimitris Kontakos for comments and suggestions, and the KNP scientific services for providing us with the data. AM acknowledges funding from the EU COST Action 'Fire Links' CA18135, Fire in the Earth System: Science & Society. Comments from two anonymous reviewers considerably improved an earlier manuscript draft.

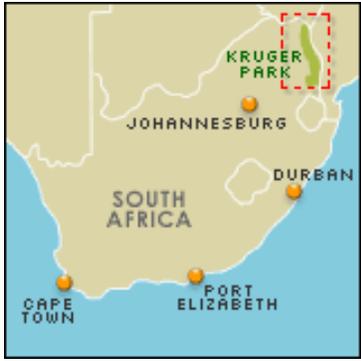a

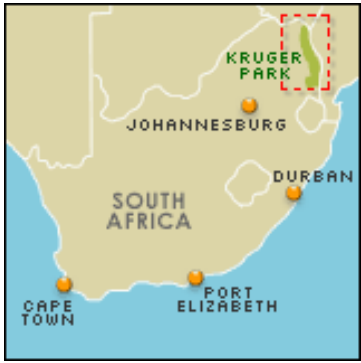a

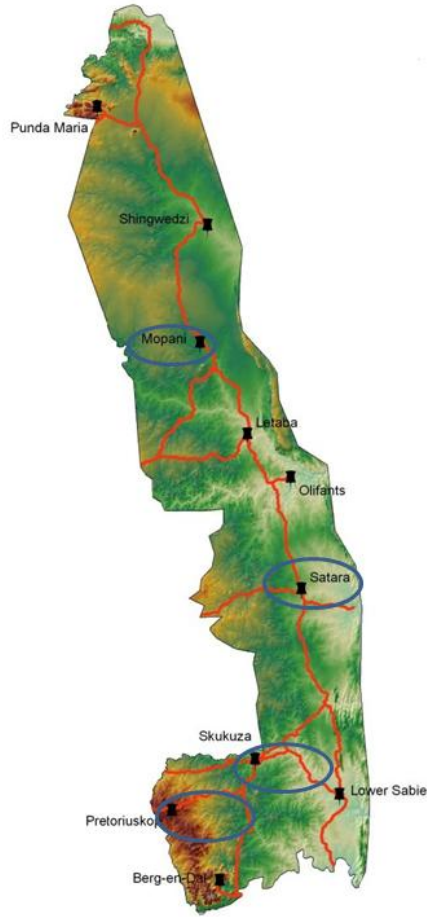

b

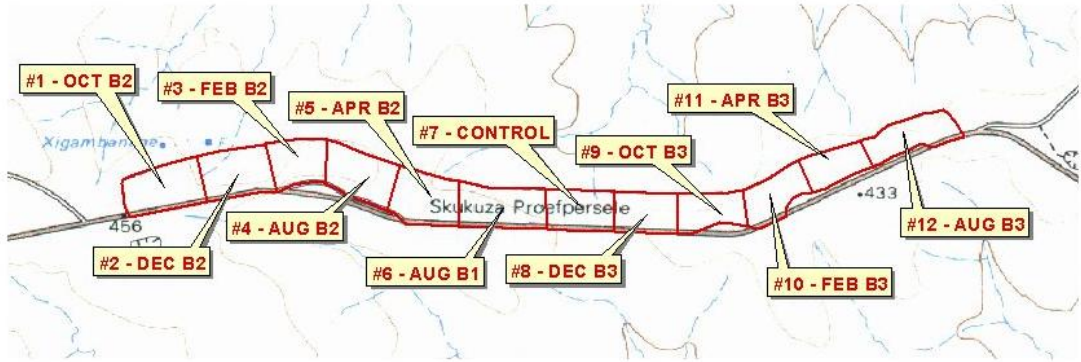

c

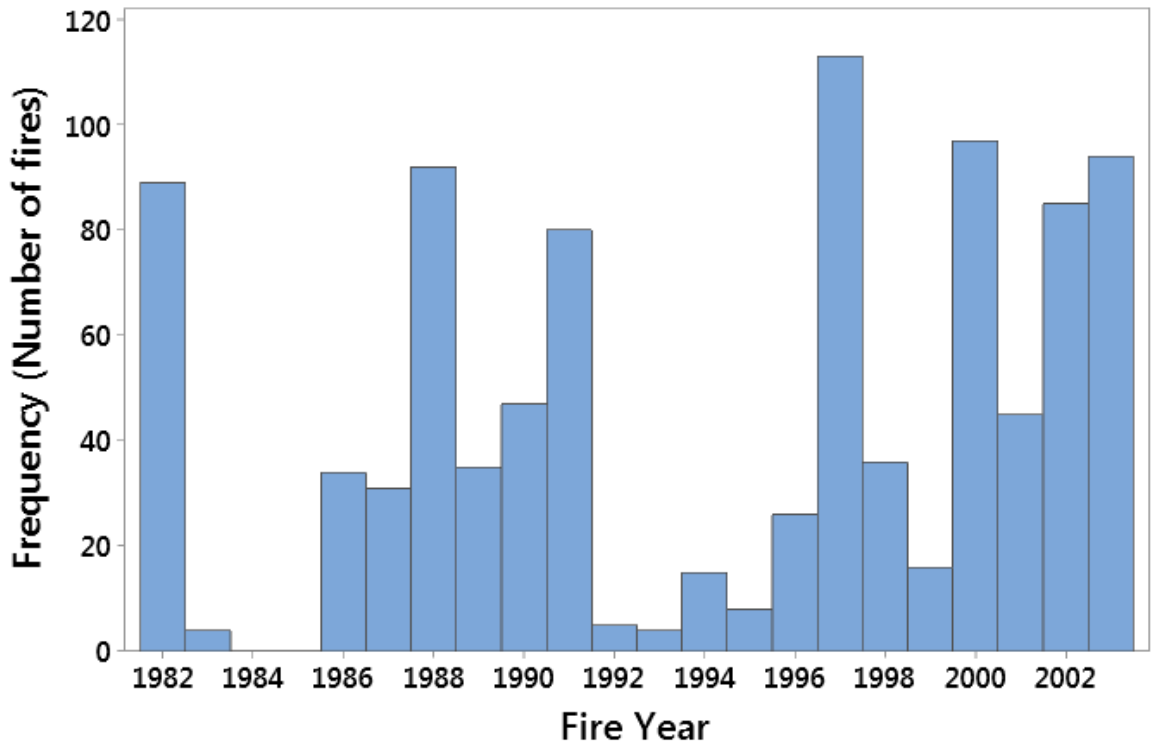

d

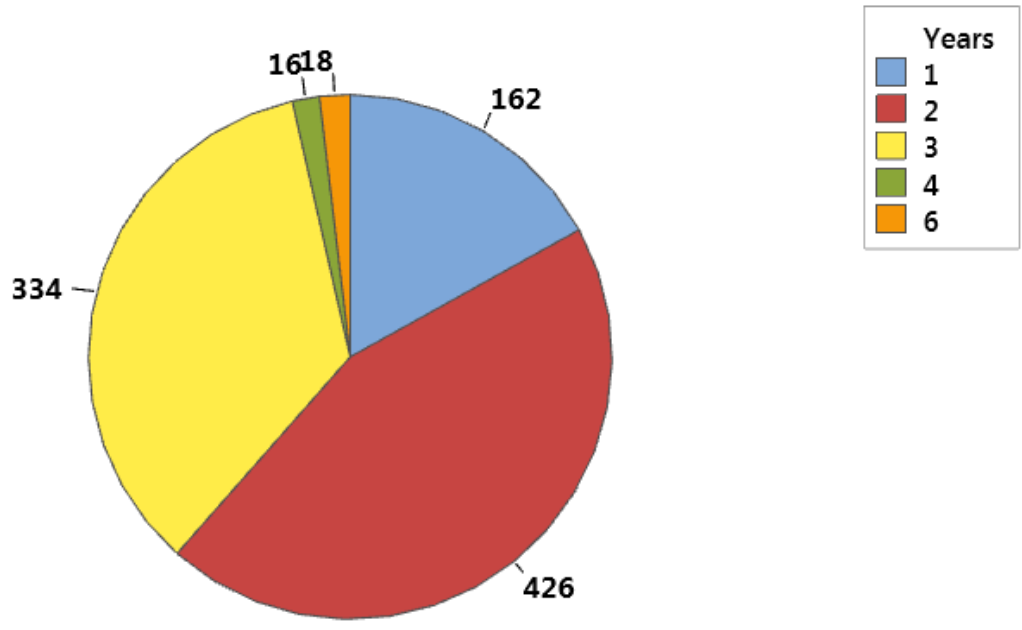
e

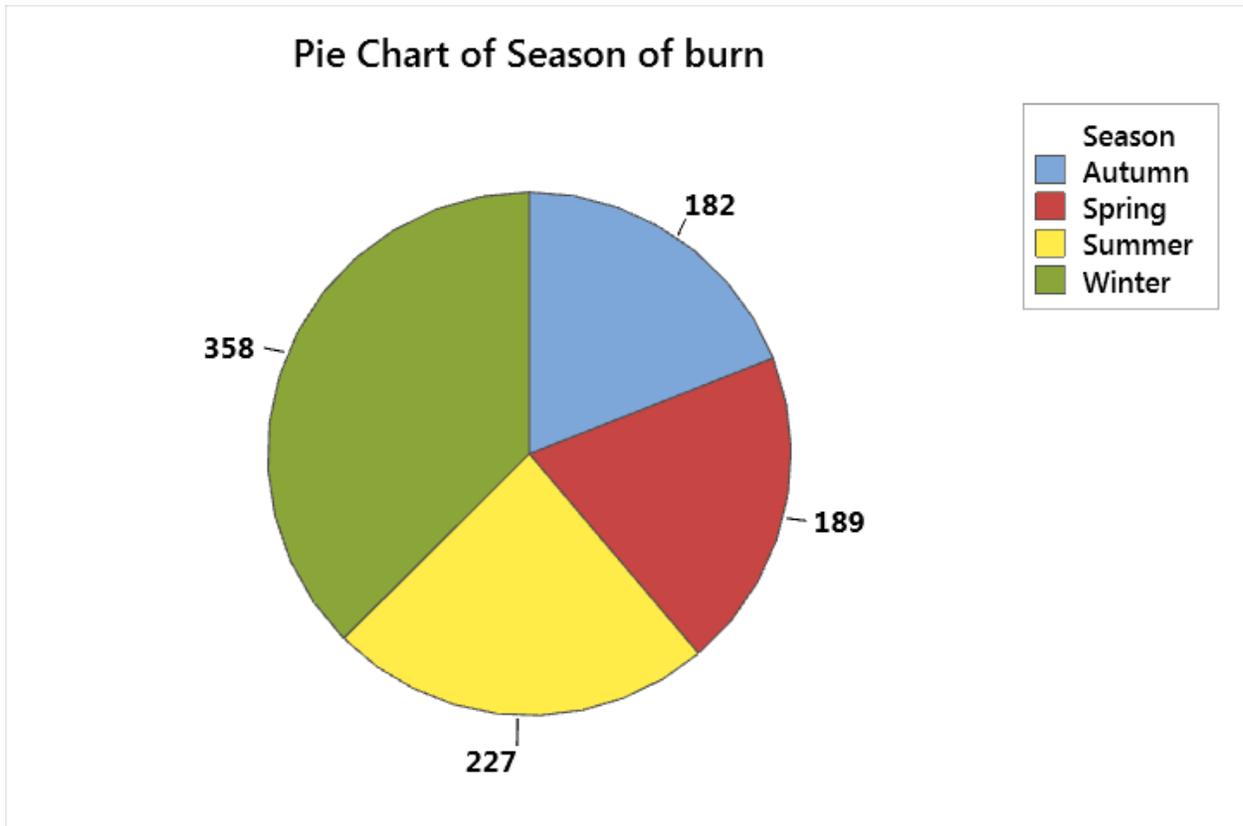

f

Figure 1. a. The location of Kruger National Park (KNP) in Southern Africa. b. A map of KNP. Deep green indicates lower elevation while dark brown higher elevation. Controlled fire burns were conducted in Experimental Burning Plots (EBP) in four savanna landscapes within the park, Pretoriuskop, Satara, Skukuza, and Mopani. In each of the four landscapes there were four EBP replicates. The four savanna landscapes are indicated with blue circles. c, Overview of an EBP. Each EBP consists of 12 fire treatments experimentally manipulating fire seasonality and fire return interval. There are 16 EBPs in total replicating all possible fire and seasonality treatment combinations in a randomized split-block design. Each treatment has a surface area of 7 ha. In the example provided the # denotes the number of treatment within the EBP, the following three letter denote the month of burn, and the letter B followed by a number indicates the fire return interval (years since last fire). For example #8 DEC B3 is the 8th treatment within this EBP, burn is always conducted in December (summer month), and the fire return interval is always every 3 years. d. Histogram of number of fires across years in the dataset. e. Pie chart of the number of fires per fire return interval (1, 2, 3, 4, and 6 years). f. Pie chart of the number of fires per season of burn (autumn, winter, spring, summer).

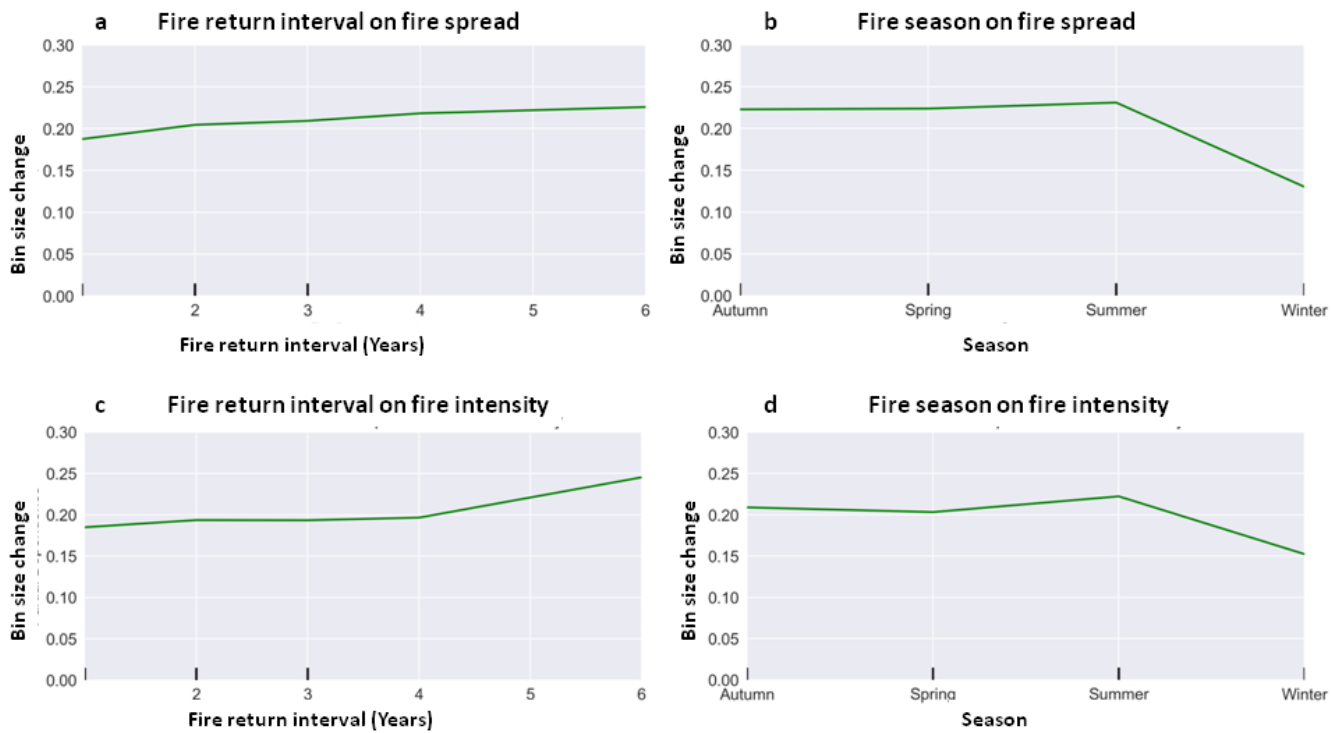

**Figure 2.** Partial dependence plots of fire return interval and seasonality on fire spread rate bin of fire intensity. A partial dependence plot indicates the chance of the current bin size of the dependent variable changing to another bin size based on the values (units) of the independent variable, all else been equal. The units of the dependent variables here are bins (of fire spread rate or intensity), with a value of 0.20 corresponding to 20% chance of bin change. . The partial effects of the first bin size of fire spread rate and first bin size of fire intensity and fire return interval or fire season are plotted here as a full plot across all bin sizes would require 20 partial plots. The full partial effects plots are provided in the Supplementary Figures 1 - 4, and there effects are very small in all cases. Results from a 75-25 % training to testing data split are shown, replicated across 10 different random states. **a.** Partial dependence of fire return interval in years on fire spread rate. **b.** Partial dependence of seasonality of fire (Fall, Winter, Spring, Autumn) on fire spread rate. **c.** Partial dependence of fire return interval in years on fire intensity. **d.** Partial dependence of fire season (Fall, Winter, Spring, Autumn) on fire intensity.

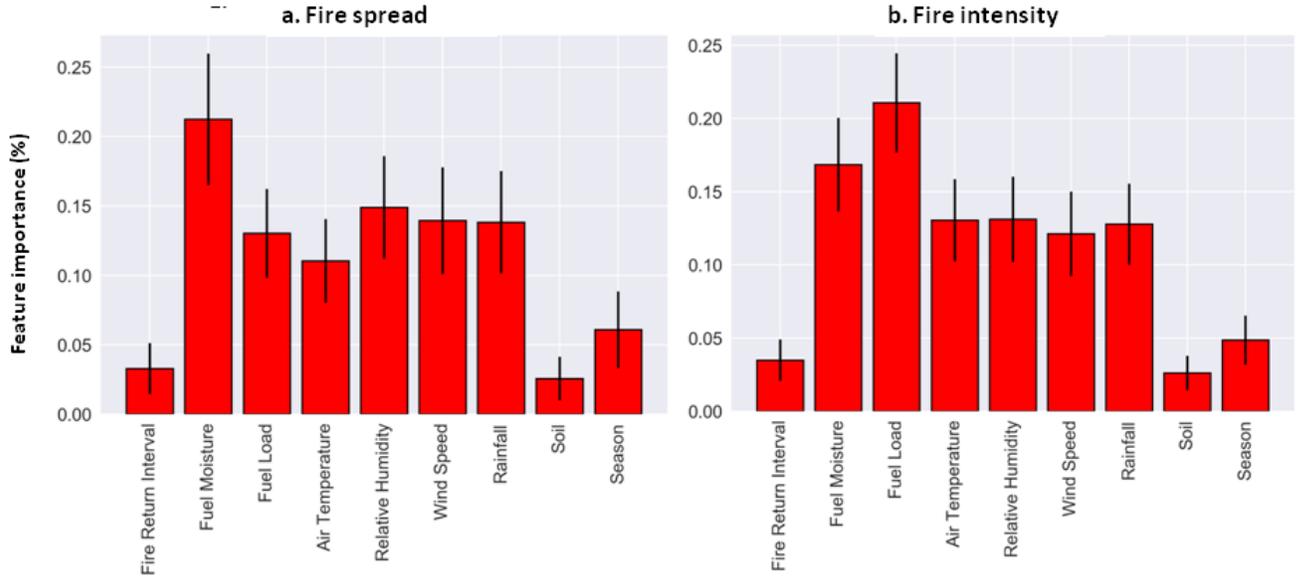

**Figure 3.** Feature importance of each independent variable, in red solid boxes. Vertical black lines indicate the 95% confidence intervals of the effect of each feature importance calculated as the standard deviation across the 10 random states. Values in each graph add to one. **a.** Feature importance of fire spread rate. Results from a 75-25% training-to-testing data split are shown, replicated across 10 different random states. **b.** Feature importance of fire intensity. Results from a 70-30% training-to-testing data split are shown, replicated across 10 different random states.

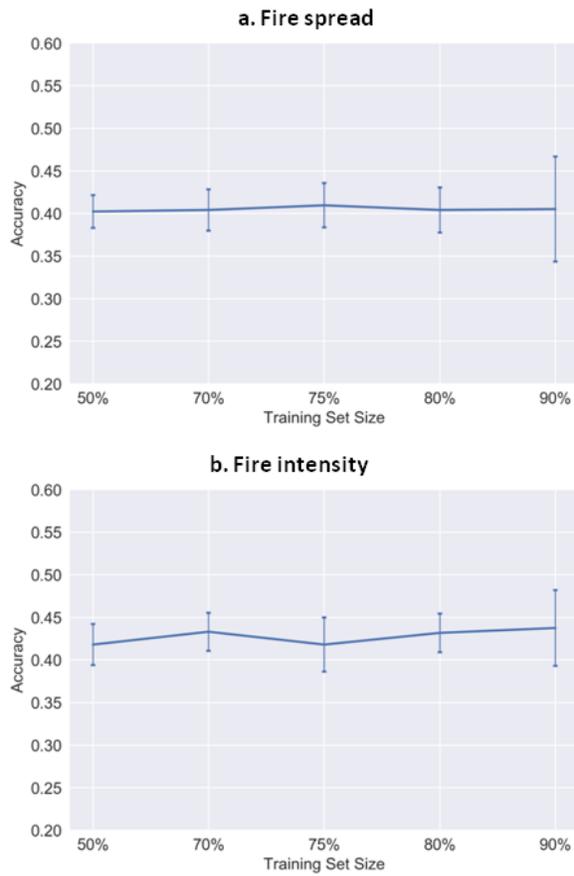

**Figure 4. a.** Mean predictive accuracy of fire spread rate based on the ratio of the training-to-testing data set size. For each dataset partitioning size, the data analysis was replicated across 10 different random states and the mean accuracy with 95% confidence intervals are plotted. **b.** Mean predictive accuracy of fire intensity based on the ratio of the training data set size as a percentage of the total data. For each dataset partitioning size the data were analyzed 10 times to account for stochasticity and mean accuracy with 95% confidence intervals are plotted.

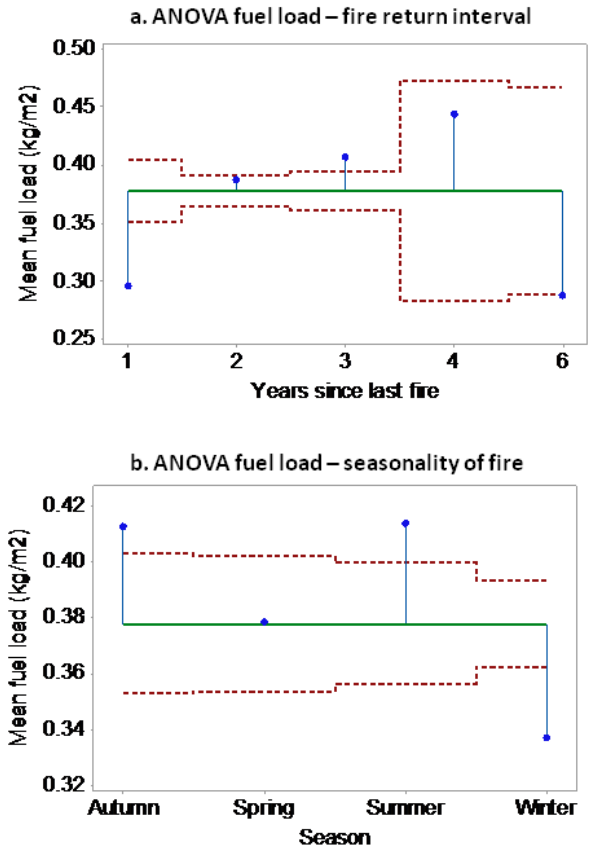

**Figure 5. a.** Analysis of variance (ANOVA) between fuel load (kg m$^{-2}$) and fire return interval (years). Horizontal solid green line indicates the mean effect of fire return interval on fuel load. Horizontal dotted red lines indicate the upper and lower 95% confidence intervals. Vertical solid blue lines indicate the effect of each fire return interval value on the grand mean. Values crossing the dotted confidence interval indicate a significant effect, while the remaining may not be differentiated from a random effect. Values crossing the upper confidence interval indicate a positive effect, while values crossing the lower indicate a negative effect on the grand mean. **b.** ANOVA between fuel load and seasonality of burn.

**Supplementary material**

# Minimal effect of prescribed burning on fire spread rate and intensity in savanna ecosystems

**Aristides Moustakas and Orestis Davlias**

**Supplementary Figures**

**Supplementary Figure 1.** Partial effect plots of fire return interval on fire spread rate across all five bin sizes

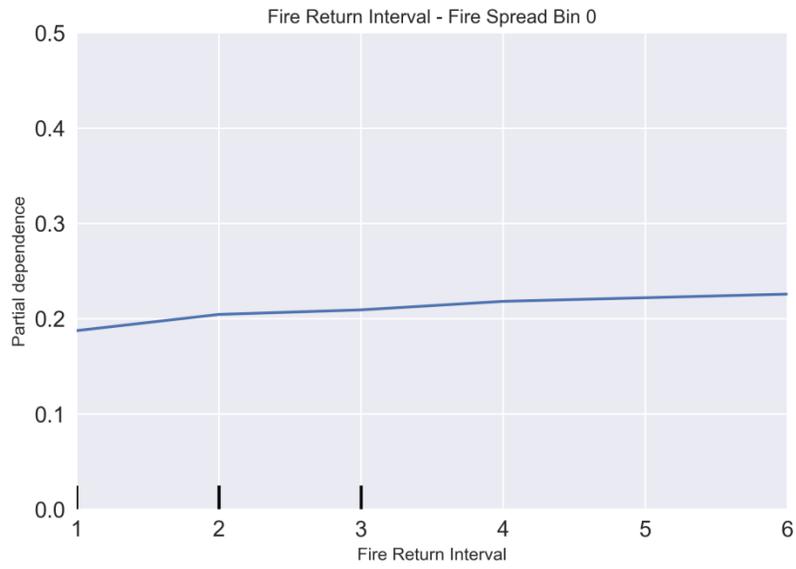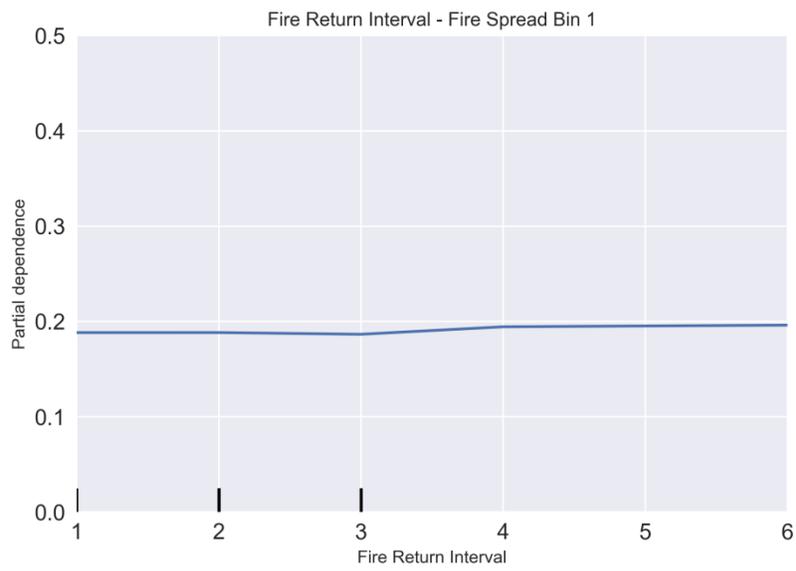

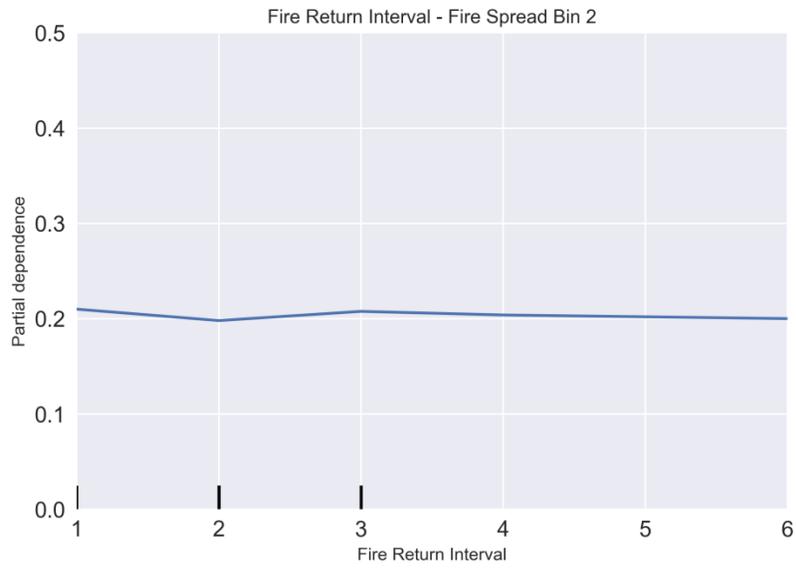

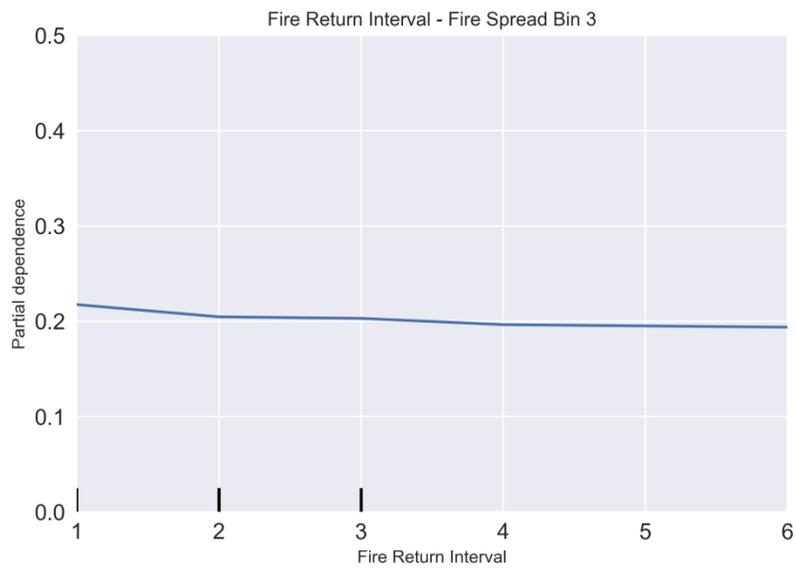

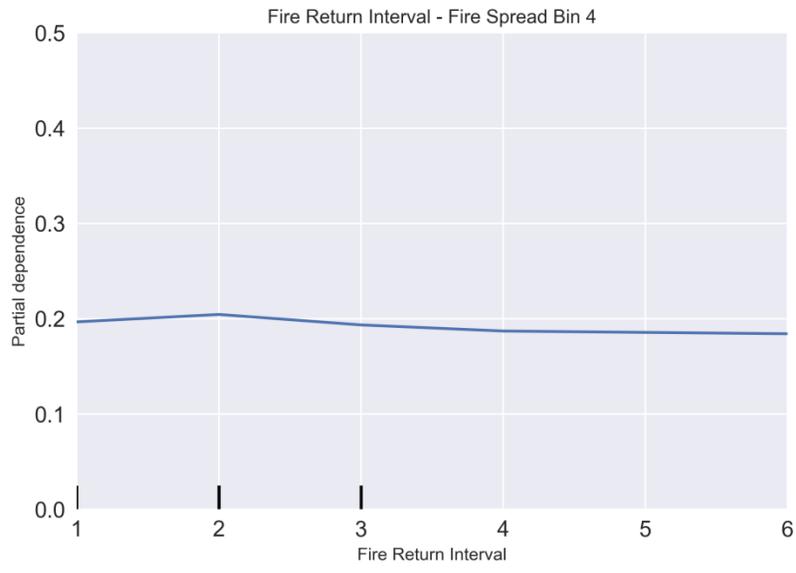

**Supplementary Figure 2.** Partial effect plots of fire season on fire spread rate across all five bin sizes

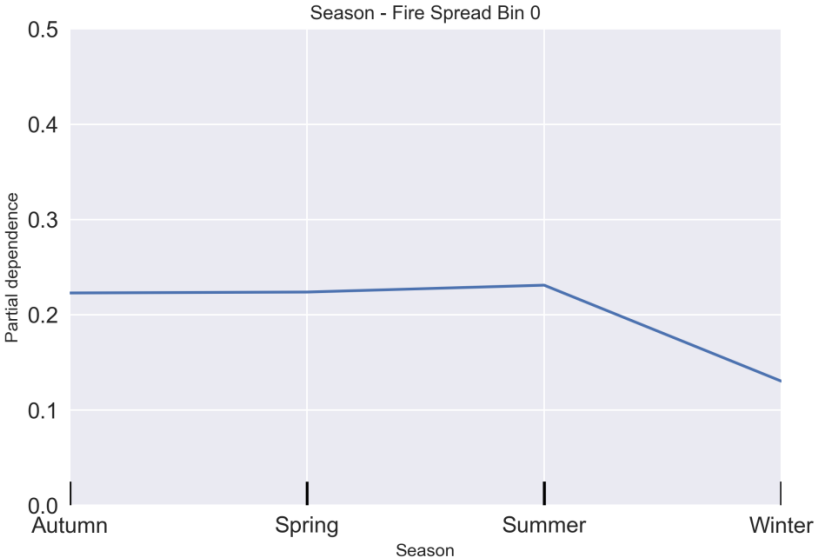

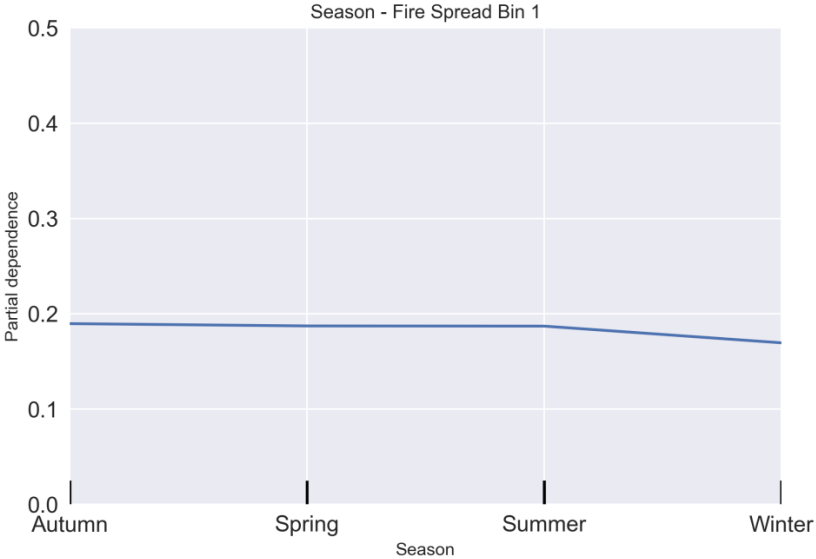

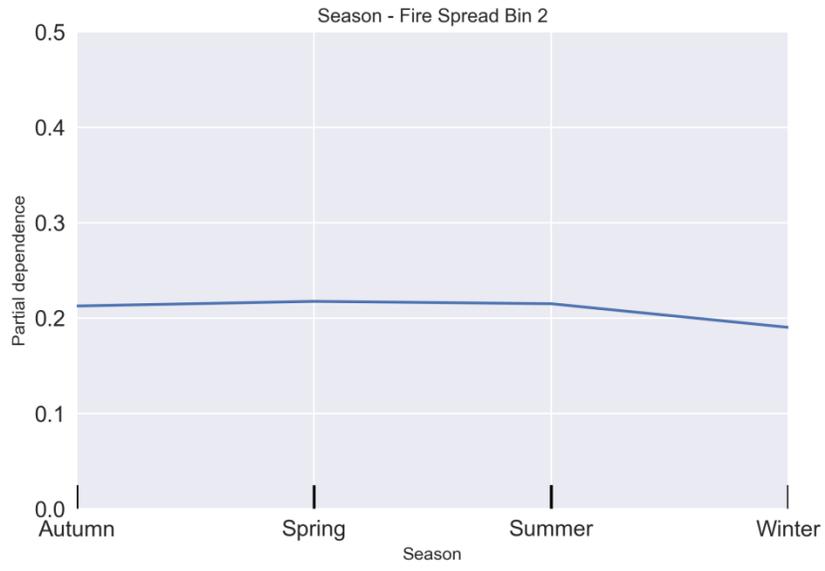
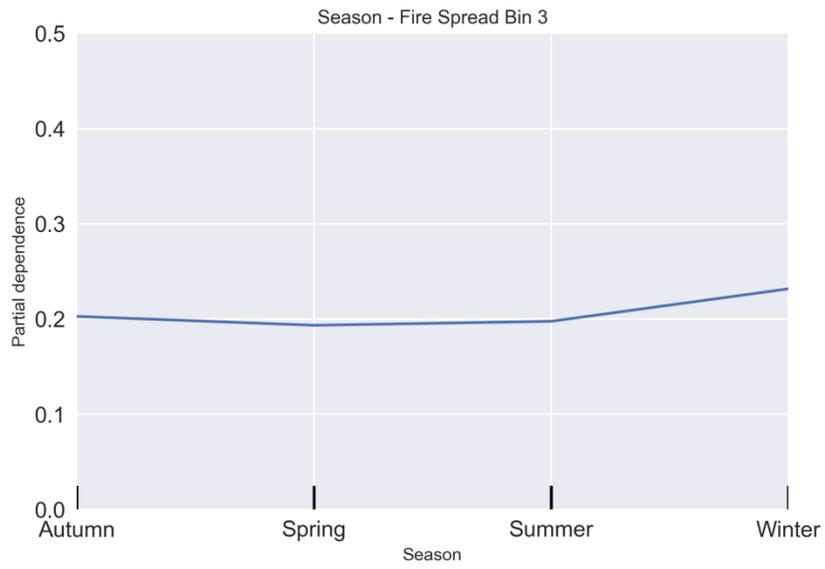

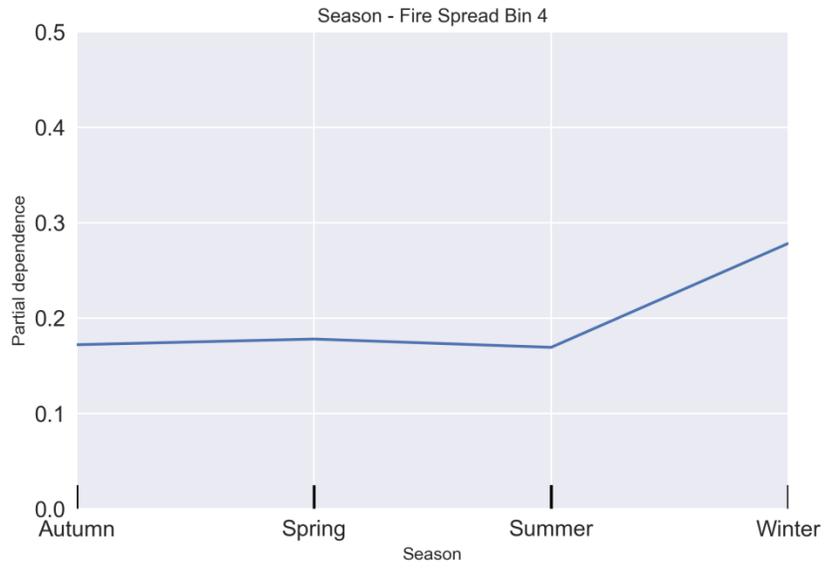

**Supplementary Figure 3.** Partial effect plots of fire return interval on fire intensity rate across all five bin sizes

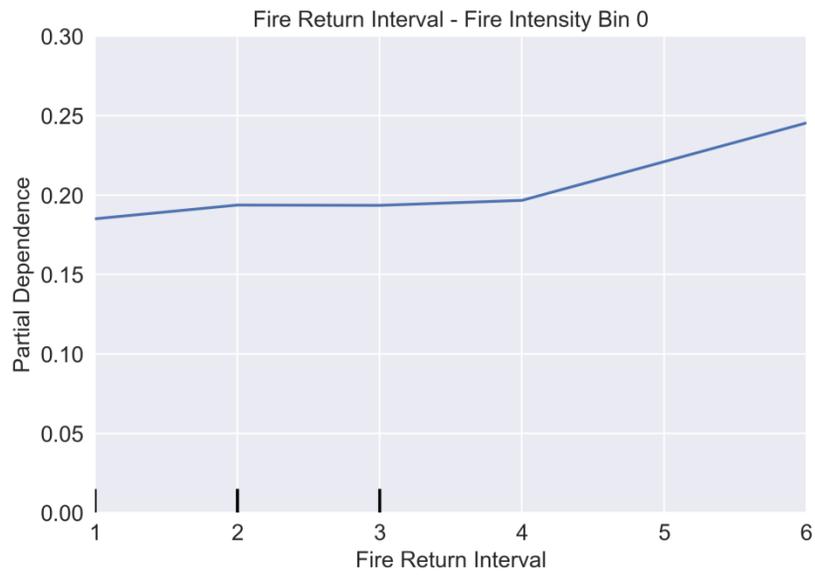

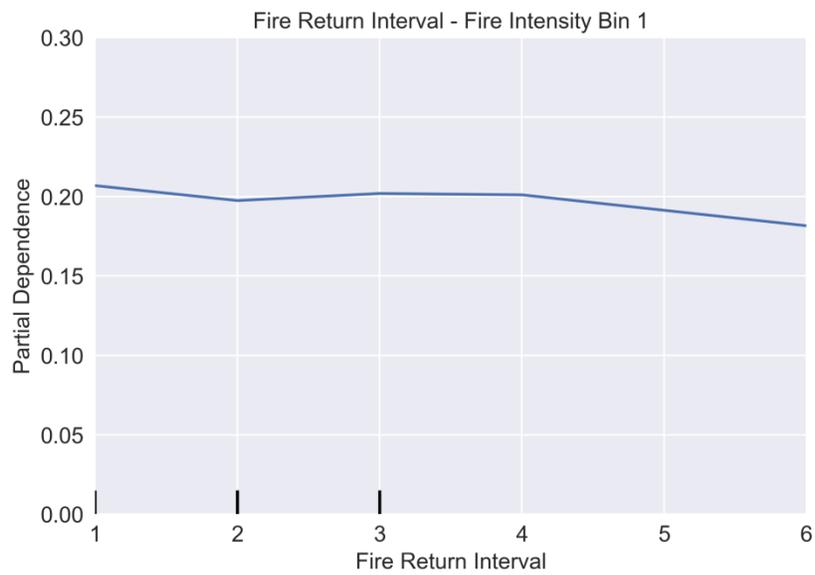

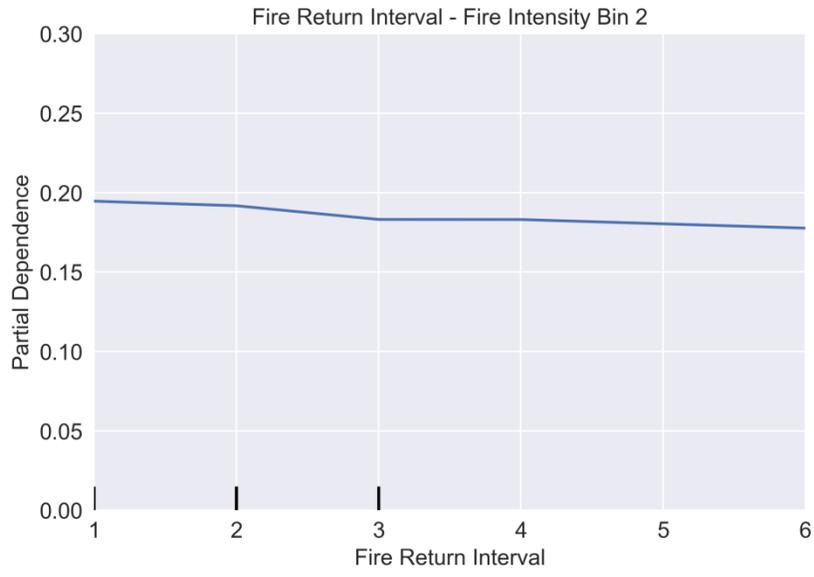

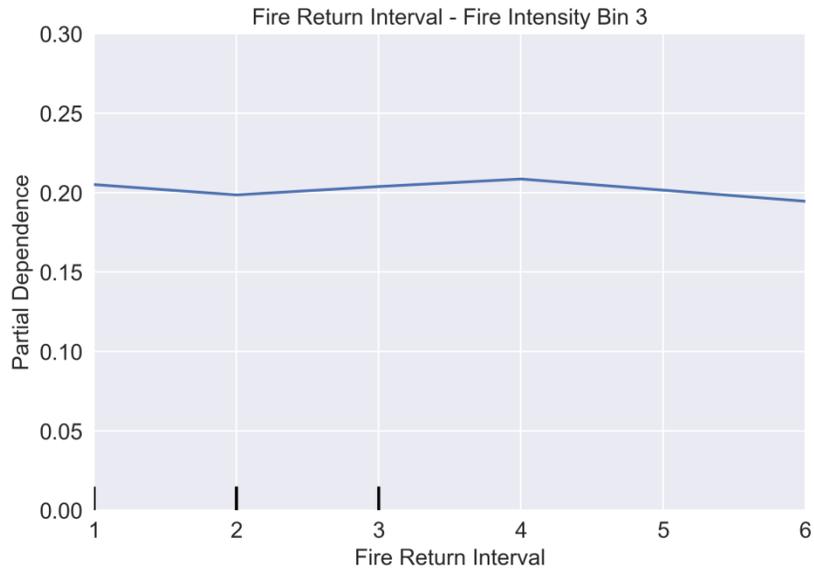

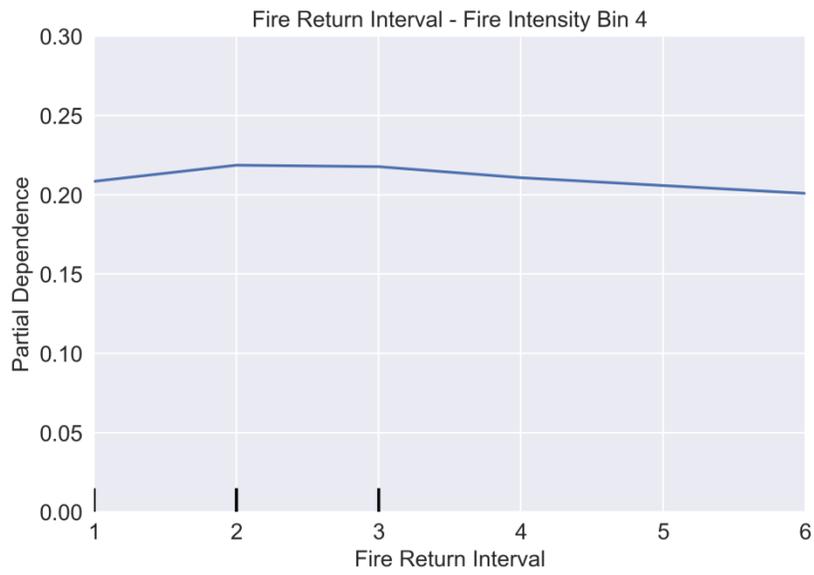

**Supplementary Figure 4.** Partial effect plots of fire season on fire intensity across all five bin sizes

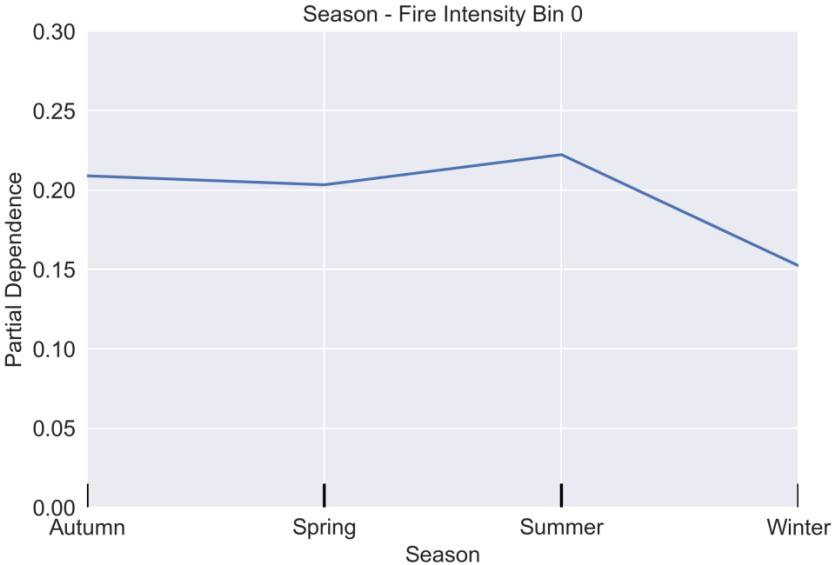

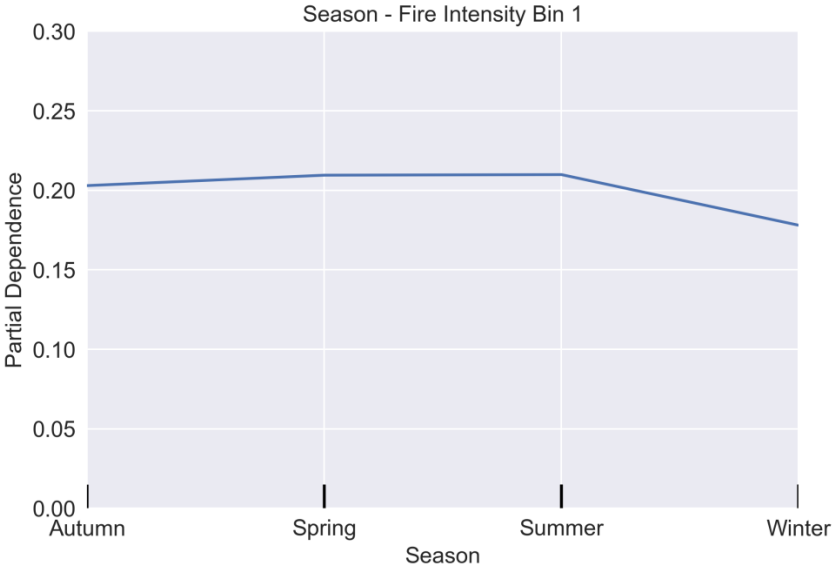

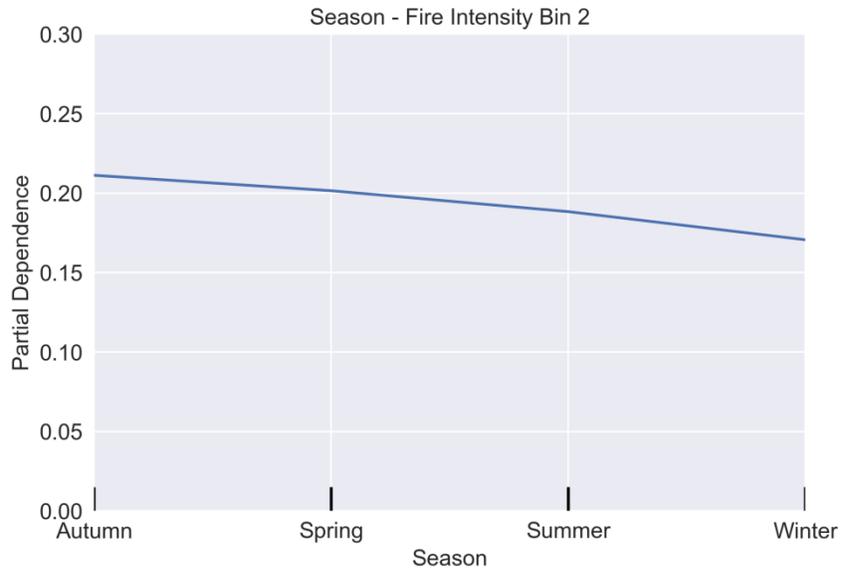

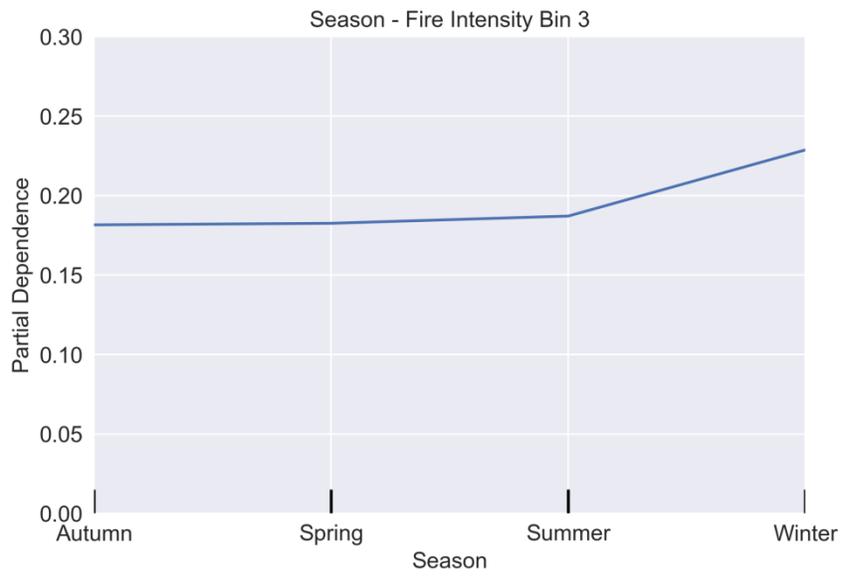

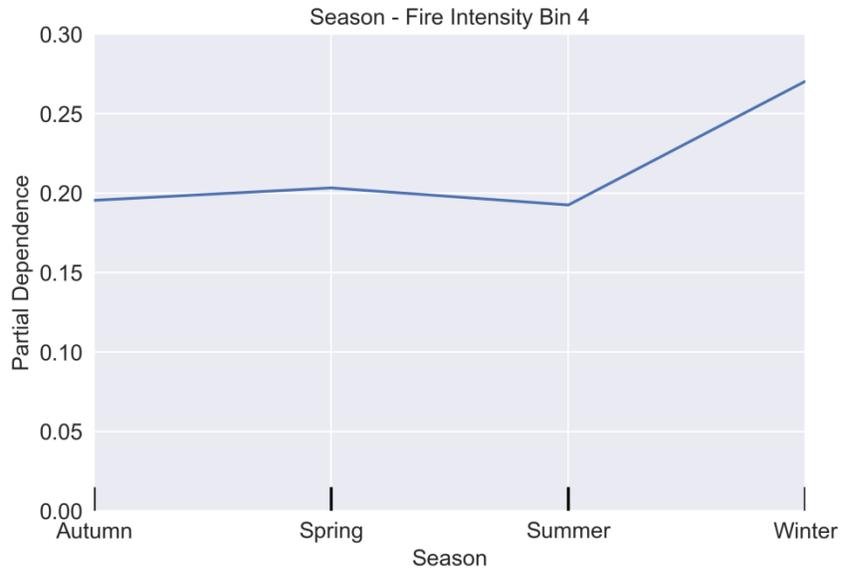

# Full code in Python

**Fire Spread Rate**
(the Python code regarding Fire intensity is identical. Simply replace 'Fire Spread' with 'Fire Intensity')

```python
# import libraries
import pandas as pd
import numpy as np
from sklearn.model_selection import train_test_split
from sklearn.preprocessing import LabelEncoder
from sklearn.ensemble import RandomForestClassifier
from sklearn.metrics import classification_report, multilabel_confusion_matrix, accuracy_score, confusion_matrix, matthews_corrcoef
from sklearn.inspection import plot_partial_dependence
from sklearn.model_selection import validation_curve
import matplotlib.pyplot as plt
import time

# import dataset
data_path = 'dataset_renamed.xls'
raw_dataset = pd.read_excel(data_path, 'Working File')
# test head
raw_dataset.head()

dataset = raw_dataset.copy()

# data pre-proccessing
# drop column MEDIAN FUEL LOAD
dataset = dataset.drop('MEDIAN FUELLOAD (kg/m2)', axis=1)
dataset = dataset.drop('Fire Intensity', axis=1)
dataset.head()

# label encode soil and season

SOIL = dataset.pop('Soil')
SEASON = dataset.pop('Season')

soil_encoder = LabelEncoder()
season_encoder = LabelEncoder()
soil_encoder.fit(SOIL)
season_encoder.fit(SEASON)
```

```python
dataset['Soil'] = soil_encoder.transform(SOIL)
dataset['Season'] = season_encoder.transform(SEASON)

# transform frequency to actual frequency
#dataset['Fire Return Interval'] = 1 / dataset['Fire Return Interval']
dataset.head()

stats = dataset.describe()
stats = stats.transpose()
stats

# sorting
dataset.sort_values(by=['Fire Spread'], inplace=True)
dataset.reset_index(drop=True, inplace=True)
print(dataset.index)
bin_1 = []
bin_2 = []
bin_3 = []
bin_4 = []
bin_5 = []

# bin stats
for i in dataset.index:
    if i < 192:
        bin_1.append(dataset['Fire Spread'][i])
    elif i < 384:
        bin_2.append(dataset['Fire Spread'][i])
    elif i < 576:
        bin_3.append(dataset['Fire Spread'][i])
    elif i < 768:
        bin_4.append(dataset['Fire Spread'][i])
    else:
        bin_5.append(dataset['Fire Spread'][i])

print('Bin #1 range:', min(bin_1), 'to', max(bin_1), 'Average:', sum(bin_1) / len(bin_1))
print('Bin #2 range:', min(bin_2), 'to', max(bin_2), 'Average:', sum(bin_2) / len(bin_2))
print('Bin #3 range:', min(bin_3), 'to', max(bin_3), 'Average:', sum(bin_3) / len(bin_3))
print('Bin #4 range:', min(bin_4), 'to', max(bin_4), 'Average:', sum(bin_4) / len(bin_4))
print('Bin #5 range:', min(bin_5), 'to', max(bin_5), 'Average:', sum(bin_5) / len(bin_5))

# binning
dataset['Fire Spread'] = dataset.index // 192

# split labels from features
```

```python
labels = np.array(dataset['Fire Spread'])
features = dataset.drop('Fire Spread', axis=1)
feature_list = list(features.columns)
features.head()

accuracy_matrix = []

for i in range(1,11):
    train_features_90, test_features_90, train_labels_90, test_labels_90 = train_test_split(features, labels, test_size = 0.1, random_state = i)
    train_features_80, test_features_80, train_labels_80, test_labels_80 = train_test_split(features, labels, test_size = 0.2, random_state = i)
    train_features_75, test_features_75, train_labels_75, test_labels_75 = train_test_split(features, labels, test_size = 0.25, random_state = i)
    train_features_70, test_features_70, train_labels_70, test_labels_70 = train_test_split(features, labels, test_size = 0.3, random_state = i)
    train_features_50, test_features_50, train_labels_50, test_labels_50 = train_test_split(features, labels, test_size = 0.5, random_state = i)

    # build and train model for 80-20
    rf_90 = RandomForestClassifier(n_estimators = 100, n_jobs = -1, random_state = i, verbose = 0)
    rf_90.fit(train_features_90, train_labels_90)

    # build and train model for 80-20
    rf_80 = RandomForestClassifier(n_estimators = 100, n_jobs = -1, random_state = i, verbose = 0)
    rf_80.fit(train_features_80, train_labels_80)

    # build and train model for 75-25
    rf_75 = RandomForestClassifier(n_estimators = 100, n_jobs = -1, random_state = i, verbose = 0)
    rf_75.fit(train_features_75, train_labels_75)

    # build and train model for 70-30
    rf_70 = RandomForestClassifier(n_estimators = 100, n_jobs = -1, random_state = i, verbose = 0)
    rf_70.fit(train_features_70, train_labels_70)

    # build and train model for 50-50
    rf_50 = RandomForestClassifier(n_estimators = 100, n_jobs = -1, random_state = i, verbose = 0)
    rf_50.fit(train_features_50, train_labels_50)

    # predict
    predictions_90 = rf_90.predict(test_features_90)
    predictions_80 = rf_80.predict(test_features_80)
    predictions_75 = rf_75.predict(test_features_75)
    predictions_70 = rf_70.predict(test_features_70)
```

```python
    predictions_50 = rf_50.predict(test_features_50)

    accuracy_matrix.append([accuracy_score(test_labels_50, predictions_50), accuracy_score(test_labels_70, predictions_70), accuracy_score(test_labels_75, predictions_75), accuracy_score(test_labels_80, predictions_80), accuracy_score(test_labels_90, predictions_90)])

accuracy_dataframe = pd.DataFrame(accuracy_matrix, columns = ['50%', '70%', '75%', '80%', '90%'])
stats = accuracy_dataframe.describe()
stats = stats.transpose()
print(stats)

means = stats['mean']
stds = stats['std']

means = stats['mean']
plt.style.use('seaborn')
fig, ax = plt.subplots(dpi = 300)
ax.set_xlabel('Training Set Size', fontsize = 14)
ax.set_ylabel('Accuracy', fontsize = 14)
xaxis = ['50%', '70%', '75%', '80%', '90%']
line_style = {"linewidth":2, "markeredgewidth":2, "elinewidth":1, "capsize":2}
acc = ax.errorbar(xaxis, means, yerr=stds, **line_style)
ax.set_ylim(0.2, 0.6)
plt.xticks(fontsize = 14)
plt.yticks(fontsize = 14)
plt.savefig('spread_accuracy', bbox_inches = 'tight')
stats.to_excel("spread_partitioning.xlsx")

rs = 42
train_features, test_features, train_labels, test_labels = train_test_split(features, labels, test_size = 0.25, random_state = rs)
base_model = RandomForestClassifier(n_jobs = -1, random_state = rs, verbose = 0)
model_fit = base_model.fit(train_features, train_labels)
model_predict = model_fit.predict(test_features)
accuracy_score(test_labels, model_predict)

# tuning n_estimators
estimators = range(100, 2000, 100)
train_score, test_score = validation_curve(RandomForestClassifier(random_state = rs),
                    X = train_features, y = train_labels,
                    param_name = 'n_estimators',
                    param_range = list(estimators), cv = 3)

fig, ax = plt.subplots(dpi = 300)
```

```python
ax.set_xlabel('Estimators', fontsize = 14)
ax.set_ylabel('Accuracy', fontsize = 14)
ax.plot(estimators, train_score)
ax.plot(estimators, test_score)
plt.savefig('spread_estimators', bbox_inches = 'tight')

estimators_dataframe = pd.DataFrame(test_score.transpose(), columns = list(estimators))
stats = estimators_dataframe.describe()
stats = stats.transpose()
fig, ax = plt.subplots(dpi = 300)
ax.set_xlabel('Estimators', fontsize = 14)
ax.set_ylabel('Mean Accuracy', fontsize = 14)
ax.plot(estimators, stats['mean'])
ax.set_xticks(list(estimators))
plt.savefig('spread_estimators_mean', bbox_inches = 'tight')

# tuning max_depth
max_depth = range(10, 120, 5)
train_score, test_score = validation_curve(RandomForestClassifier(n_estimators = 100),
                    X = train_features, y = train_labels,
                    param_name = 'max_depth',
                    param_range = list(max_depth), cv = 3)

fig, ax = plt.subplots(dpi = 300)
ax.set_xlabel('Max Depth', fontsize = 14)
ax.set_ylabel('Accuracy', fontsize = 14)
ax.plot(max_depth, train_score)
ax.plot(max_depth, test_score)
plt.savefig('spread_max_depth', bbox_inches = 'tight')

test_score.transpose()
max_depth_dataframe = pd.DataFrame(test_score.transpose(), columns = list(max_depth))
stats = max_depth_dataframe.describe()
stats = stats.transpose()
fig, ax = plt.subplots(dpi = 300)
ax.set_xlabel('Max Depth', fontsize = 14)
ax.set_ylabel('Mean Accuracy', fontsize = 14)
ax.plot(max_depth, stats['mean'])
ax.set_xticks(list(max_depth))
plt.savefig('spread_max_depth_mean', bbox_inches = 'tight')

# tuning min_sample_split
min_samples_split = range(2, 10, 1)
train_score, test_score = validation_curve(RandomForestClassifier(n_estimators = 100, max_depth = 30),
```

```python
                              X = train_features, y = train_labels,
                              param_name = 'min_samples_split',
                              param_range = list(min_samples_split), cv = 3)

fig, ax = plt.subplots(dpi = 300)
ax.set_xlabel('Minimum Samples to split', fontsize = 14)
ax.set_ylabel('Accuracy', fontsize = 14)
ax.plot(min_samples_split, train_score)
ax.plot(min_samples_split, test_score)
plt.savefig('spread_min_samples_split', bbox_inches = 'tight')

test_score.transpose()
min_samples_split_dataframe = pd.DataFrame(test_score.transpose(), columns = list(min_samples_split))
stats = min_samples_split_dataframe.describe()
stats = stats.transpose()
fig, ax = plt.subplots(dpi = 300)
ax.set_xlabel('Minimum Samples to split', fontsize = 14)
ax.set_ylabel('Mean Accuracy', fontsize = 14)
ax.plot(min_samples_split, stats['mean'])
ax.set_xticks(list(min_samples_split))
plt.savefig('spread_min_samples_split_mean', bbox_inches = 'tight')

# tuning min_sample_leaf
min_samples_leaf = range(2, 10, 1)
train_score, test_score = validation_curve(RandomForestClassifier(n_estimators = 100, max_depth = 30,
min_samples_split = 7),
                              X = train_features, y = train_labels,
                              param_name = 'min_samples_leaf',
                              param_range = list(min_samples_leaf), cv = 3)

fig, ax = plt.subplots(dpi = 300)
ax.set_xlabel('Minimum min_samples_leaf', fontsize = 14)
ax.set_ylabel('Accuracy', fontsize = 14)
ax.plot(min_samples_leaf, train_score)
ax.plot(min_samples_leaf, test_score)
plt.savefig('spread_min_samples_leaf', bbox_inches = 'tight')

test_score.transpose()
min_samples_leaf_dataframe = pd.DataFrame(test_score.transpose(), columns = list(min_samples_leaf))
stats = min_samples_leaf_dataframe.describe()
stats = stats.transpose()
fig, ax = plt.subplots(dpi = 300)
ax.set_xlabel('min_samples_leaf', fontsize = 14)
ax.set_ylabel('Mean Accuracy', fontsize = 14)
```

```python
ax.plot(min_samples_leaf, stats['mean'])
ax.set_xticks(list(min_samples_leaf))
plt.savefig('spread_min_samples_leaf_mean', bbox_inches = 'tight')

rf = RandomForestClassifier(n_estimators = 100, min_samples_split=7, min_samples_leaf=4, max_depth=30, n_jobs = -1, random_state = rs, verbose = 1)
rf.fit(train_features, train_labels)

# variable importances
importances = list(rf.feature_importances_)
std = np.std([tree.feature_importances_ for tree in rf.estimators_],
         axis=0)
feature_importances = [(feature, round(importance, 2)) for feature, importance in zip(feature_list, importances)]
feature_importances = sorted(feature_importances, key = lambda x: x[1], reverse = True)
[print('Variable: {:20} Importance: {}'.format(*pair)) for pair in feature_importances]

x_values = list(range(len(importances)))
plt.rcParams['figure.dpi'] = 300
plt.style.use('seaborn')
plt.bar(x_values, importances, orientation = 'vertical', color = 'red', edgecolor = 'k', linewidth = 1.2, yerr = std)
plt.xticks(x_values, feature_list, rotation = 'vertical', fontsize = 14)
plt.yticks(fontsize = 14)
plt.title('Variable Importances')
plt.savefig('spread_importances', bbox_inches = 'tight')

predictions = rf.predict(test_features)

print(multilabel_confusion_matrix(test_labels, predictions, labels=[0, 1, 2, 3, 4]))
print(classification_report(test_labels,predictions))
print(accuracy_score(test_labels, predictions))

features = [8]
fig = plot_partial_dependence(rf, train_features, features, target = 0)
plt.rcParams['figure.dpi'] = 300
plt.xlabel('Season')
plt.axis([0, 3, 0, 0.5])
sn = [0, 1, 2, 3]
xn = range(len(sn))
sn = season_encoder.inverse_transform(sn)
plt.xticks(xn, sn, fontsize=14)
plt.yticks(fontsize = 14)
plt.title('Season Partial Dependance on Fire Spread')
plt.savefig('spread_season', bbox_inches = 'tight')
```

```
features = [0]
fig = plot_partial_dependence(rf, train_features, features, target = 0)
plt.rcParams['figure.dpi'] = 300
plt.xlabel('Fire Return Interval')
plt.axis([1, 6, 0, 0.5])
plt.xticks(fontsize=14)
plt.yticks(fontsize = 14)
plt.title('Fire Return Interval Partial Dependance on Fire Spread')
plt.savefig('spread_interval', bbox_inches = 'tight')

features = [1]
fig = plot_partial_dependence(rf, train_features, features, target = 0)
plt.rcParams['figure.dpi'] = 300
plt.xlabel('Fuel Moisture')
plt.axis([10, 150, 0, 0.50])
plt.xticks(fontsize=14)
plt.yticks(fontsize = 14)
plt.title('Fuel Moisture Partial Dependance on Fire Spread')
plt.savefig('spread_moisture', bbox_inches = 'tight')

features = [2]
fig = plot_partial_dependence(rf, train_features, features, target = 0)
plt.rcParams['figure.dpi'] = 300
plt.xlabel('Fuel Load')
plt.axis([1500, 6500, 0, 0.50])
plt.xticks(fontsize=14)
plt.yticks(fontsize = 14)
plt.title('Fuel Load Partial Dependance on Fire Spread')
plt.savefig('spread_load', bbox_inches = 'tight')

features = [3]
fig = plot_partial_dependence(rf, train_features, features, target = 0)
plt.rcParams['figure.dpi'] = 300
plt.xlabel('Air Temperature')
plt.axis([20, 36, 0, 0.50])
plt.xticks(fontsize=14)
plt.yticks(fontsize = 14)
plt.title('Air Temperature Partial Dependance on Fire Spread')
plt.savefig('spread_temperature', bbox_inches = 'tight')

features = [4]
fig = plot_partial_dependence(rf, train_features, features, target = 0)
plt.rcParams['figure.dpi'] = 300
plt.xlabel('Relative Humidity')
```

```
plt.axis([24, 76, 0, 0.50])
plt.xticks(fontsize=14)
plt.yticks(fontsize = 14)
plt.title('Relative Humidity Partial Dependance on Fire Spread')
plt.savefig('spread_humidity', bbox_inches = 'tight')

features = [5]
fig = plot_partial_dependence(rf, train_features, features, target = 0)
plt.rcParams['figure.dpi'] = 300
plt.xlabel('Wind Speed')
plt.axis([0.6, 3.6, 0, 0.50])
plt.xticks(fontsize=14)
plt.yticks(fontsize = 14)
plt.title('Wind Speed Partial Dependance on Fire Spread')
plt.savefig('spread_speed', bbox_inches = 'tight')

features = [6]
fig = plot_partial_dependence(rf, train_features, features, target = 0)
plt.rcParams['figure.dpi'] = 300
plt.xlabel('Rainfall')
plt.axis([264, 1272, 0, 0.50])
plt.xticks(fontsize=14)
plt.yticks(fontsize = 14)
plt.title('Rainfall Partial Dependance on Fire Spread')
plt.savefig('spread_rainfall', bbox_inches = 'tight')

features = [7]
fig = plot_partial_dependence(rf, train_features, features, target = 0)
plt.rcParams['figure.dpi'] = 300
plt.xlabel('Soil')
plt.axis([0, 1, 0, 0.50])
sn = [0, 1]
xn = range(len(sn))
sn = season_encoder.inverse_transform(sn)
plt.xticks(xn, sn, fontsize=14)
plt.yticks(fontsize = 14)
plt.title('Soil Partial Dependance on Fire Spread')
plt.savefig('spread_soil', bbox_inches = 'tight')
```